\documentstyle[12pt,aaspp4,rotating]{article}
\input{psfig.sty}
\def\lax    {${_<\atop^{\sim}}$}
\def\gax    {${_>\atop^{\sim}}$}

\def\etal   {{\it et al.}~}
\def\pg     {PG1114+445}
\def\Lya    {Ly$\alpha$}

\begin{document}
\title{Discovery of Associated Absorption Lines in an X-ray Warm
Absorber: HST$^1$ Observations of PG1114+445 }
\author{Smita Mathur$^2$, Belinda Wilkes, Martin Elvis}
\affil{Harvard-Smithsonian Center for Astrophysics,
60 Garden St., Cambridge, MA 02138}
\footnotetext[1]{Based on observations with the NASA/ESA {\it Hubble Space
Telescope},
obtained at the Space Telescope Science Institute, which is operated
by the Association of Universities for Research in Astronomy, Inc.,
under NASA contract NAS5-26555.}
\footnotetext[2]{smita@cfa.harvard.edu}

\setcounter{footnote}{1}

\received{}
\accepted{}
\lefthead{Mathur \etal~}
\righthead{Absorption Lines in PG1114+445}

\section*{Abstract}

The unified picture of X-ray/UV absorbers offers an unique
opportunity to probe the nuclear environment of AGNs. To test the
unified absorber scenario and to understand the physical properties of
the absorber we obtained the first UV spectrum of
\pg\, using HST FOS. \pg\ is known to have an X-ray ionized
(``warm'') absorber, so that UV absorption is predicted.  The HST
spectrum clearly shows strong UV absorption lines due to \Lya,
CIV and NV, blueshifted by $\sim$600~km~s$^{-1}$ with respect to
the quasar redshift. Since both X-ray and UV absorbers are rare in
radio-quiet quasars, these observations argue strongly that
the X-ray and UV absorbers are
closely physically related,  probably identical. We place constraints
on the parameters of the absorber and conclude  that the
mass outflow rate is comparable to the accretion rate in \pg. ~

\newpage
\section{Introduction}

The discovery that the OVII, OVIII X-ray and OVI UV absorption in
 quasar 3C351 could be readily modeled by a single a X-ray/UV
absorber (Mathur \etal 1994) offered new and unique way to determine
the physical conditions of the absorbing material in  AGN.
As more  X-ray/UV absorbers were found
\footnote{The `red' quasar 3C212 (Mathur 1994); a variable
Seyfert galaxy NGC5548 (Mathur, Elvis \& Wilkes 1995 ); the broad
absorption line quasars (BALQSO) PHL~5200 (Mathur, Elvis \& Singh
1995); the Seyfert galaxy
NGC~3516, explaining the previous presence and present
disappearance of its broad absorption lines (Mathur {\it
et~al}~1997); and the Seyfert galaxy NGC 3783 (Shields
 \& Hamann 1997).}
it became clear that, although they span a wide range of parameter
space, they are systematically: highly ionized; with high column density; low
density; outflowing; and situated outside the broad emission line
region. With these properties the X/UV absorbers delineate a
previously unrecognized
nuclear component of AGNs with important consequences - a wind
or outflow which carries significant kinetic energy and a mass loss rate
comparable to the accretion rate needed to power the AGN
continuum (Mathur, Elvis \& Wilkes 1995: MEW95). The X-ray/UV connection was
further strengthened when Crenshaw
 (1997) found that UV and X-ray absorption tend to occur
together in Seyfert 1 galaxies.
It is  possible that a wide range of associated
absorbers including Broad Absorption Line systems, are related through a
continuum of properties such as
the column density, ionization parameter, and outflow velocity.

First though we
need to establish how common the X/UV absorbers are; whether the X-ray
absorbers are really physically related to the UV absorbers in all AGN;
whether this unified picture is quite general
or is merely a chance coincidence in a minority of cases. In order
to answer these questions, as well as to probe the nuclear material in
more objects,
 we initiated a HST program to search quasars known to have X-ray ionized
(``warm'') absorbers for the UV
absorbers predicted by the unified X-ray/UV absorber model. Though
common in Seyfert galaxies and radio-loud quasars, UV absorption is
rare in radio-quiet quasars ($\S 3.1$); hence these observations give
 a statistically strong test.

Here we present a positive detection for the first object observed:
PG1114+445. This is a radio-quiet quasar at z=0.144 and was the
only one of a well defined sub-sample of 23 PG quasars to show an
ionized absorber signature
in its ROSAT spectrum (Laor \etal\ 1994).

\section{The HST Spectra}

On 23 November 1996 we observed \pg\ with the Faint Object
Spectrograph (FOS) on board the Hubble Space Telescope (HST),
using a 1$^{\prime\prime}$ aperture. The source acquisition was
done in a four step sequence to ensure pointing accuracy of
0.12$^{\prime\prime}$. Grating G130H was used on the blue side of
the detector and G190H and G270H on the red side. The total
exposure was 12,820 seconds on G130, 4,360 seconds on G190 and
180 seconds on G270H. We will focus only on the properties of the
absorption lines in
this paper. The analysis and discussion of the emission line
properties is deferred to another paper (Wills \etal~1998).

The data were reduced with IRAF using the standard procedure
described in the HST data handbook.  The data accumulated with
different exposures were combined and wavelength calibrated. The
final spectra
are shown in Figure~1. Strong absorption lines are clearly observed
superimposed on the emission lines of \Lya, NV, and CIV, as
expected from the X/UV models. The absorption lines are much deeper
than the continuum level implying that the broad emission line region
(BELR) is partially covered by the absorbing material.  The models also predict
OVI absorption
lines. The G130 spectrum around OVI (observed $\lambda=1183$ \AA ~)
 has very low signal-to-noise, but does show a marginally
detected  OVI
absorption line. Absorption
in the low ionization
line of MgII is not detected, again as expected given the high
ionization of the X-ray absorber.

 The measurement of
absorption lines situated within a broad emission line profile is
inherently uncertain since the real shape of the emission line is
unknown. We use two separate methods to measure the line equivalent
widths (EW). The
initial analysis of the absorption lines used the IRAF task {\bf
splot} and the results are given in Table 1. We estimate the {\bf
splot} errors on the EW of
the \Lya\ and CIV lines to be $\sim$40\%. The errors on the NV
doublet are smaller, $\sim$10\%, because the NV emission line is
very weak and so the continuum for the absorption lines is much
better defined. The CIV absorption doublet is highly blended
 for two reasons: (1) the resolution of
G190H is not as high (1.47 \AA\ per diode compared to 1 \AA\ per
diode of G130H); (2) the separation between the components of the
CIV doublet is smaller (2.57 \AA\ compared to 4 \AA\ for
NV). As a result  it is
difficult to measure the EWs of the individual components and Table
1  lists the total EW of the CIV absorption
doublet. The NV doublet components have about the same EW, with
a doublet ratio of about 1 indicating that the lines are saturated.
The troughs do not reach zero intensity; higher resolution
observations would be required to detect the dark cores of the
lines. It is also possible that the line cores do not reach zero
intensity as a result of scattering and/or partial covering
(Korista \etal\ 1992, Barlow \etal\ 1997). The present observations,
however, do not have high enough resolution to detect such
effects. The OVI absorption line is only marginally detected, so we have
estimated only a rough lower limit on the EW.

Fitting the emission and absorption lines together, in some cases
allows better estimates of the EWs of the lines.  Accordingly
the {\bf stsdas} task {\bf specfit} (contributed by G. Kriss) was
also used to determine line parameters, subject to some assumptions on
the shape of the emission line profile.  We characterized the continuum with a
simple power-law, the emission lines with multiple Gaussians, and
 each absorption line with a single Gaussian. We first fitted the
continuum in a featureless part of the spectrum. The parameters
of the power-law continuum, the slope and normalization, were
then fixed in subsequent fits. The parameters of the emission lines
(flux, centroid, FWHM and skew) and absorption lines (EW,
centroid and FWHM) were allowed to be free. The fits to the \Lya,
NV and CIV profiles are shown in Figure 2 and the corresponding EW and
FWHM of the
absorption lines are given in Table 1. The two methods, {\bf splot}
and {\bf specfit} give consistent results, although the {\bf specfit}
results are systematically higher. These uncertainties do not
dominate the current discussion.

The FWHM of the lines is
\gax$500$ km s$^{-1}$, well resolved given the nominal spectral resolution of
1--1.5 \AA\ ($\sim 200$ km s$^{-1}$) (The FWHM of the Milky Way
interstellar MgII lines is 2--2.2 \AA,
consistent with the nominal resolution of 2.09 \AA\ in G270H) . This
would imply that the
absorber is dispersed in
velocity space. The redshift of the absorption lines is 0.142 (Table
1), blueshifted
by  600~km~s$^{-1}$ compared to the quasar redshift of
0.144. The redshift of OVI line could not be measured accurately due
to its very low S/N, but may be slightly offset to the blue. All the
parameters of OVI given in Table 1 are highly uncertain.

\section {The X/UV absorber}

The HST spectrum of \pg\ shows the associated high ionization
\Lya, NV, CIV (and possibly OVI)
absorption lines predicted  from the X/UV models. Is this just a
chance coincidence?

About 10\% of Seyfert galaxies observed with IUE have associated
absorption lines
in their UV spectra (Ulrich, 1988),
so there is a 10\% chance of finding them in any
randomly selected Seyfert galaxy. In the HST sample of Seyfert
galaxies Crenshaw (1997) finds the fraction
 of $\sim$ 50\%. However,  for higher luminosity quasars,
strong narrow associated absorption  appears to arise
predominantly in steep spectrum, radio-loud objects (Foltz \etal~1988)
rather than in radio-quiet quasars such as \pg.~  In fact, there has been some
evidence of a dichotomy between the occurrence of narrow associated
absorption in radio-loud quasars, and broad associated absorption
in radio-quiet quasars (i.e. in BALQSOs, Foltz {\it et
al.}~1988). Only 1 of 29 radio-quiet quasars showed associated
absorption in the sample studied by Foltz \etal\

Similarly, in X-rays, warm absorbers are observed in about half
of the Seyfert 1 galaxies (Reynolds, 1997). But
absorption in radio-quiet quasars is not as frequent.
In the complete sample of 20 radio-quiet PG quasars observed by
Laor \etal\
(1994), \pg~ was the only object with evidence for
absorption by ionized matter.

The probability for \pg\ to have associated UV absorption {\it and}
X-ray ionized absorption by chance is then $\sim$1.7$\times 10^{-3}$. The
two  absorption
systems must be physically related.

\subsection {Physical Properties of the Ionized Gas}
A detailed
comparison of the UV and X-ray absorbers is required to
see if they originate in the same physical component of the nuclear
material of the AGN.

The X-ray properties of the ionized absorber in \pg, based on
ASCA observations, are given in George \etal\ (1997). The
absorber showed an oxygen edge with optical depth of about 3.1 at
0.76 keV, reminiscent of the oxygen edges seen in the ROSAT
spectra of 3C351 (Fiore \etal 1993) and NGC3783 (Turner \etal
1993).  The total equivalent hydrogen column density of the \pg\
absorber is $\sim
2\times 10^{22}$ cm$^{-2}$ and the X-ray ionization parameter
$U_x\sim 0.1$ (corresponding to a traditional ionization parameter
U=16.67)
\footnote{U is defined as the dimensionless ratio of ionizing photon
 to hydrogen number density. Photons with energy larger than 1 Rydberg
 are integrated. In U$_x$ only X-ray photons between 0.1 keV and 10 keV
 are integrated (Netzer 1996)}. Using the best fit warm absorber model, George
 \etal~ predicted  the
  column densities of the relevant ions using the photoionization code ION
 (Table~1). The shape of the input continuum used in the
 ION code is defined in Netzer (1996). The shape of the input continuum
 can affect the ionization state of the gas considerably (MEW95). So
 we also used two other ionizing continua as input
 to the photoionization code CLOUDY (Ferland 1996) to derive the
 column densities
 of the UV-observed ions.
If we use the standard AGN
 continuum given in
CLOUDY (corresponding to U=5.37), the derived column densities are 2-5 times
larger (Table~1). The observed continuum of \pg~ itself is
somewhat different again. In the  CLOUDY AGN continuum  the X-ray
power-law of slope ($f_{\nu}\propto \nu^{-\alpha}$) $\alpha=$0.7 extends down
 to 0.3 keV, while below 0.3 keV there is
a steep upturn into the EUV region. The observed X-ray power-law
index in \pg~ is 0.8 all the way down to 0.11 keV.
The observed continuum (corresponding to U=6.46) gives larger UV line  column
 densities
 (Table 1) by  a factor of 3-9 compared to the CLOUDY AGN continuum.
Clearly, the ionic column densities are sensitive to the
shape of the ionizing continuum. Since the important EUV range is
unobservable,  we cannot derive the column densities accurately.

To compare the UV and X-ray absorbers we now have to convert
the EWs of the absorption lines in the HST spectra  to
the column densities of the ions. However, as discussed in $\S~2$,
the lines are saturated and so column densities cannot be
measured directly without knowing the velocity dispersion
parameter (\underline{b} parameter) of the lines. The \underline{b}
parameter must be less than the FWHM of the lines (Table 1), so b\lax$ 500$ km
s$^{-1}$. We  obtain  lower limits to the column densities
assuming them to be on the linear portion of the
curve-of-growth  using both {\bf splot} and {\bf specfit}
values (Table~1).  We
find these column density limits to be  consistent with those
predicted from the
X-ray data (Table~1). For the  UV and X-ray values to match we further
require
\underline{b}\gax$ 300$ km s$^{-1}$ (see MEW95 for the details of the
method). This constrains \underline{b} to  300\lax \underline{b}
\lax 500 km s$^{-1}$. Higher resolution observations can test this
directly. Moreover, given this allowed \underline{b} range  the UV
absorption lines cannot split into more
than 2--3 components if all correspond to the X-ray absorber.

The ready quantitative agreement of the UV and X-ray derived ion
column densities argues that the UV and X-ray absorption is
highly likely to originate in the same material. At a minimum, the X-ray
absorber makes a large contribution to the absorption seen in the UV.

{}From the blueshifts of the absorption lines the outflow velocity
of the X/UV absorber is about 600~km~s$^{-1}$.
Outflows are
normal in associated absorption systems and can be used to derive
some interesting numbers (MEW95) assuming the
UV and X-ray absorbers are the same. We know that the  absorber has
column density
of N$_H=2\times 10^{22}$ atoms cm$^{-2}$ and ionization parameter U=5.37
(for the CLOUDY AGN continuum). From photoionization modeling
the distance of the absorber from the nucleus is r=2.4$\times10^{18}n_5^{-1/2}$
{}~cm (where n$_5$ is density in the units of 10$^{5}$ cm$^{-3}$).
The depth of the absorption lines is clearly larger than the continuum
level (figure 1), so the absorber must be situated outside the broad
emission line region (BELR), similar to other X/UV absorbers
($\S~1$). Scaling from the BELR size from the reverberation mapping of
NGC5548 (Clavel \etal~1991) for \pg~ luminosity
(L$_{Bol}=5\times10^{45}$ erg s$^{-1}$) yields a distance of
the BELR from the central continuum of $\sim 1/10$ pc. So the distance
of the absorber r\gax $3.7\times 10^{17}$~cm. This
puts an upper limit on the density, n$<43 n_5$ cm$^{-3}$. The mass
of the absorber is then $120 f_{0.1} n_5^{-1}$~M$_{\odot}$ for a covering
factor of f=10\%. The line of sight mass outflow rate would be
\.{M}$_{out}$=1.1$f_{0.1}$ M$_{\odot}$ yr$^{-1}$ comparable to the
accretion rate of 0.9 M$_{\odot}$ yr$^{-1}$ needed to power \pg~
 at 10\% efficiency. The
line of sight kinetic energy carried out in the flow is not as large, however,
$\sim10^{41}$ erg s$^{-1}$ (c.f \.{M}$_{out}$=10$f_{0.1}$ M$_{\odot}$
yr$^{-1}$ in NGC5548 and kinetic energy of $\sim 10^{43}$ erg~s$^{-1}$,
L$_{Bol}=5\times10^{44}$ erg s$^{-1}$).

\section{Conclusions}

The HST FOS UV spectrum of \pg~ contains associated high
ionization, UV absorption lines, as predicted from models of the
X-ray ionized absorber.  We find that the absorber is situated outside
the BELR and
outflowing with a line of sight velocity of $\sim600$ km s$^{-1}$. The mass
outflow rate of 1.1 M$_{\odot}$~yr$^{-1}$ for a 10\% covering factor,
is comparable to the accretion rate onto the nuclear black-hole.

The rarity of both UV and X-ray absorbers
individually in radio-quiet quasars virtually requires that the X-ray
and UV absorbers are closely physically related.  The close similarity
of the column
densities obtained from both UV and X-ray data  strongly
suggests that they are identical. Thus the absorber in \pg\ satisfies
both statistical as well as physical tests of our X/UV absorber model.

We would like to thank Ian George for useful discussions and for an
early copy of his paper. Ian Evans is thanked for discussion regarding
the FOS calibration.
SM is supported through NASA grants NAG5-3249
(LTSA) and GO-06484.01-95A from Space Telescope Science Institute. BJW
is supported through NASA contract NAS8-39073 (ASC).


\newpage

\noindent
{\bf Figure Captions:}\\

\noindent
{\bf Figure 1:} The HST spectrum of \pg\ : absorption lines of \Lya, NV and
CIV are clearly seen. The expected position of OVI absorption line is
marked by a vertical bar. The G270H spectrum is smoothened by two
pixels. The lower line in each plot is the error spectrum.

\noindent
{\bf Figure 2:} {\bf specfit} fit to the emission and absorption lines (a)
\Lya\ and NV doublet (b) CIV.  \\

\newpage
\thispagestyle{empty}

\begin{table}[h]
\vspace*{8.5in}
\begin{rotate}{90}
{\bf Table 1:} ~Absorption Line Parameters
\end{rotate}
\hspace*{2.0in}
\begin{rotate}{90}
\begin{tabular}{|llcllllllll|}
\hline\hline
 &  & z &\multicolumn{2}{c}{EW (\AA)} &
	\multicolumn{2}{c|}{FWHM(km~s$^{-1}$)}&Observed$^b$ N$_{ion}$&
		\multicolumn{3}{|c|}{Predicted N$_{ion}$} \\
Line,$\lambda$  & $\lambda_{obs}$ (\AA) &  &
	{\bf splot}$^a$ & {\bf specfit} & {\bf splot} &
		{\bf specfit}&&ION$^c$&CLOUDY$^d$&SED$^e$\\
\hline
&&&&&&&&&& \\
\Lya, 1215.7& 1389.1& 0.1426$\pm 0.0001$ &2.2& 3.6 $\pm 0.5$ & 500&
 812$\pm$103&
	$>$3-6$\times$10$^{14}$ & &1.0$\times$10$^{16}$&1.2$\times$10$^{16}$\\
&&&&&&&&&& \\
CIV$^f$, 1549.1 & 1769.5& 0.1423$\pm 0.0003$ &5.2& 5.9$\pm 0.1$& 980&
1422$\pm$79&

$>$7-9$\times$10$^{14}$&$3.4\times$10$^{14}$&1.9$\times$10$^{15}$&7.0$\times$10$^{15}$\\
&&&&&&&&&& \\
NV, 1238.8 & 1415.7& 0.1428$\pm 0.0004$ &2.1& 3.1$\pm 0.2$ & 470& 573$\pm$43&

$>$8-13$\times$10$^{14}$&$3.9\times$10$^{15}$&9.6$\times$10$^{15}$&8.9$\times$10$^{16}$ \\
{}~ ~ ~ 1242.8 & 1420.2& 0.1426$\pm 0.0002$ &2.2& 2.7$\pm 0.2$ & 500&
634$\pm$48&&&& \\
&&&&&&&&&& \\
OVI$^g$, 1033.8   & 1179  & 0.14 & 2  &          &  430  &      &

$>$1$\times$10$^{15}$&4.5$\times$10$^{17}$&6.1$\times$10$^{17}$&3.8$\times$10$^{18}$ \\
&&&&&&&&&& \\
&&&&&&&&&& \\
\hline
\hline
\end{tabular}
\end{rotate}

\smallskip
\small
\noindent
\newline
\hspace*{4.0in}
\begin{rotate}{90}
a. error estimates are 40\% on \Lya ~and CIV, 10\% on NV\\
\end{rotate}
\hspace*{0.3in}
\begin{rotate}{90}
b. Two numbers correspond to {\bf splot} and {\bf specfit} values.\\
\end{rotate}
\hspace*{0.3in}
\begin{rotate}{90}
c. Input continuum as in photoionization code ION, from George \etal\
1997 \\
\end{rotate}
\hspace*{0.3in}
\begin{rotate}{90}
d. Input continuum as ``AGN'' in photoionization code CLOUDY\\
\end{rotate}
\hspace*{0.3in}
\begin{rotate}{90}
e. CLOUDY AGN continuum modified to observed X-ray parameters.\\
\end{rotate}
\hspace*{0.3in}
\begin{rotate}{90}
f. Measurements apply to both lines in the doublet combined. \\
\end{rotate}
\hspace*{0.3in}
\begin{rotate}{90}
g. OVI absorption line  only marginally detected, measurements
highly uncertain.\\
\end{rotate}
\newline

\end{table}


\newpage
\begin{figure}
\psfig{figure=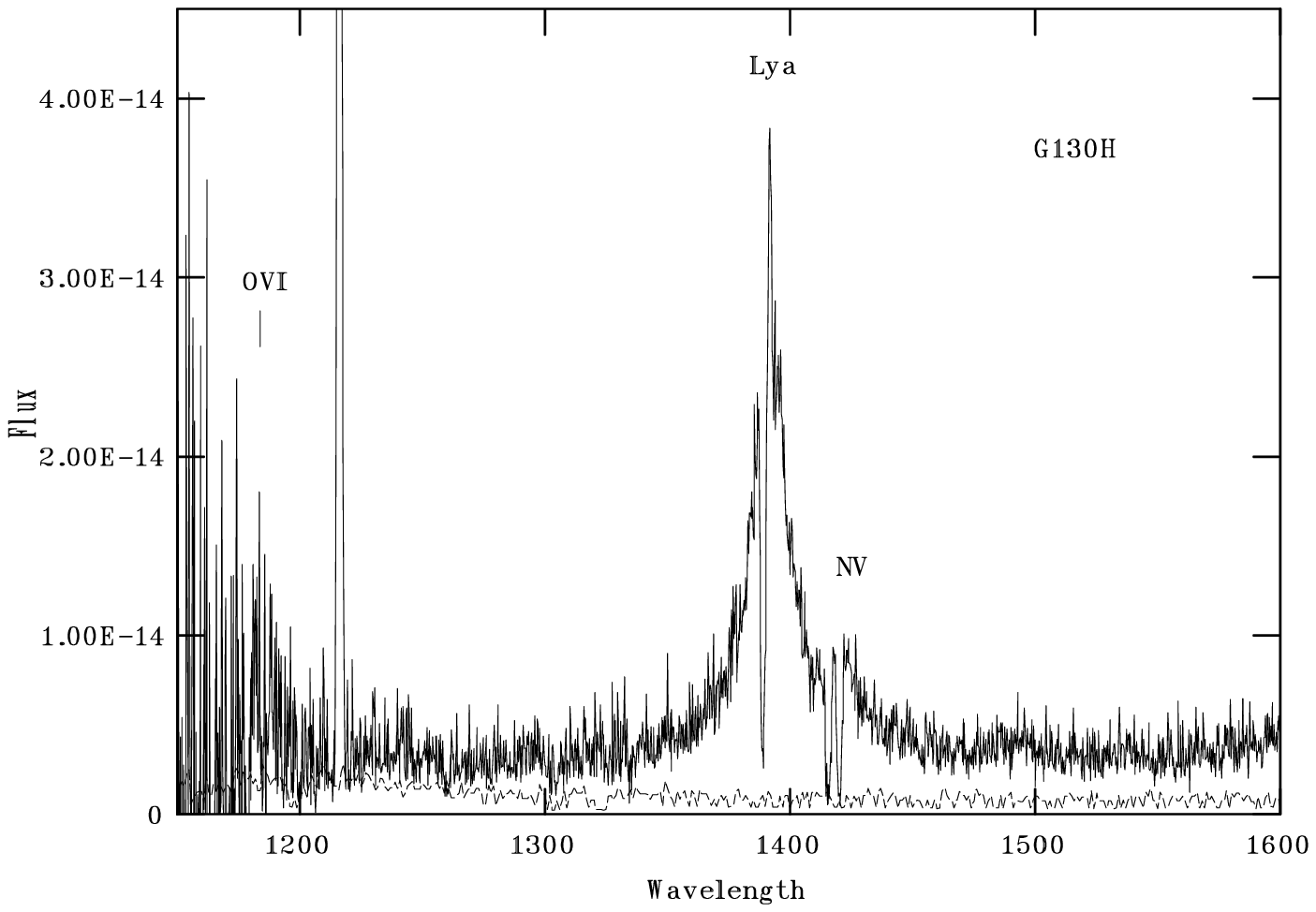,height=2.0truein,width=8.0truein}
\vspace*{1in}
\psfig{figure=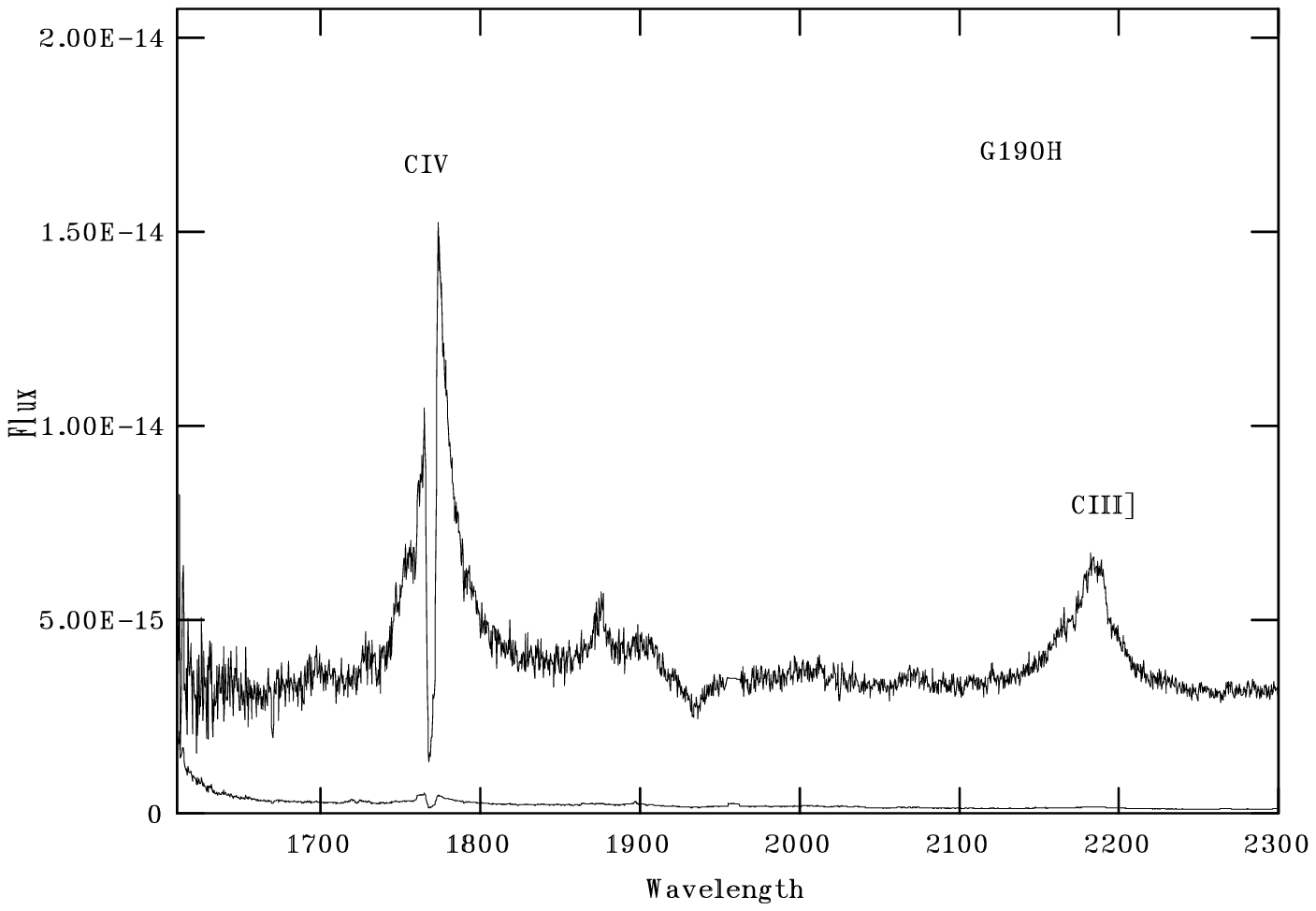,height=2.0truein,width=8.0truein}
\vspace*{1in}
\psfig{figure=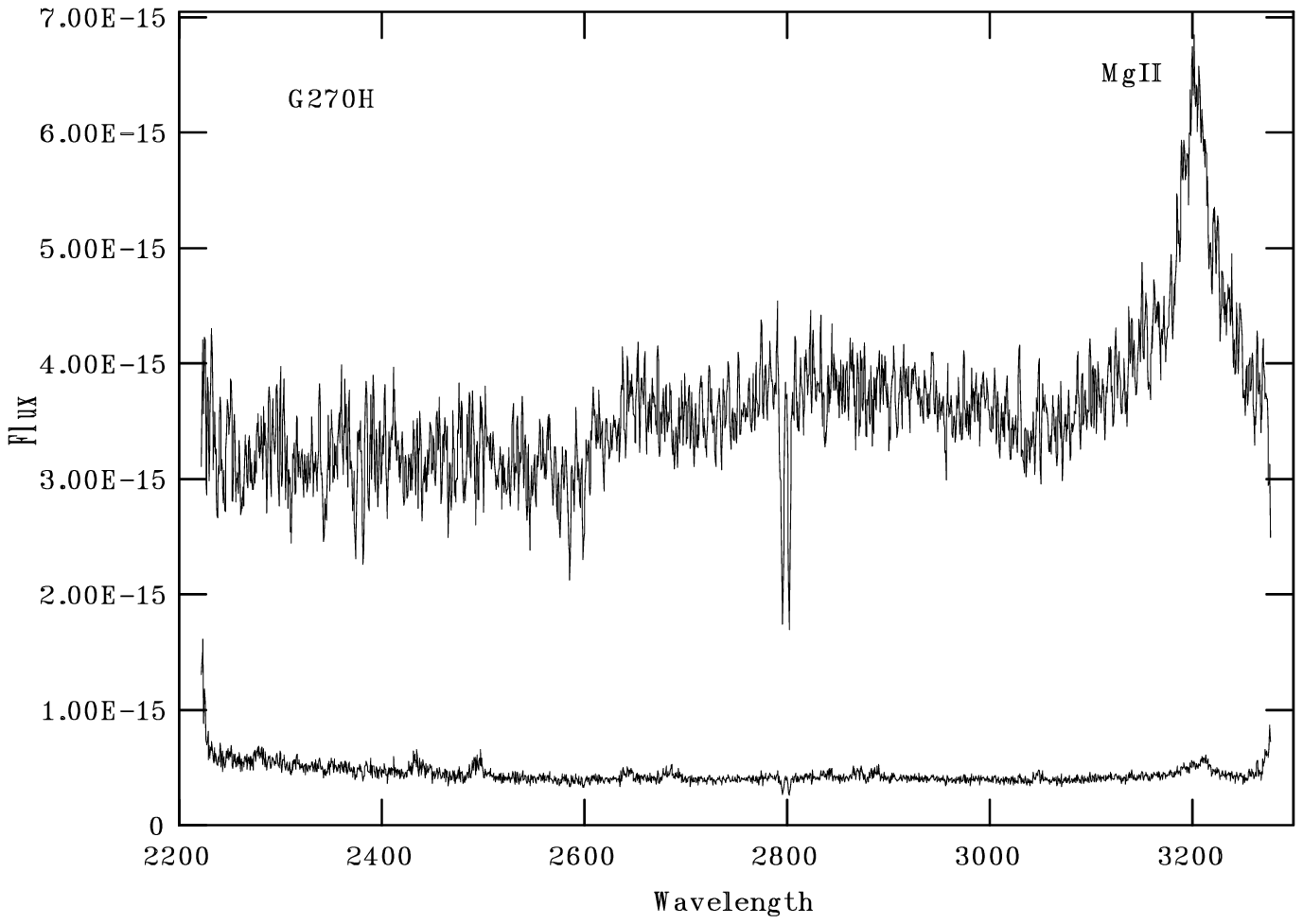,height=2.0truein,width=8.0truein}
\end{figure}

\newpage
\begin{figure}
\psfig{figure=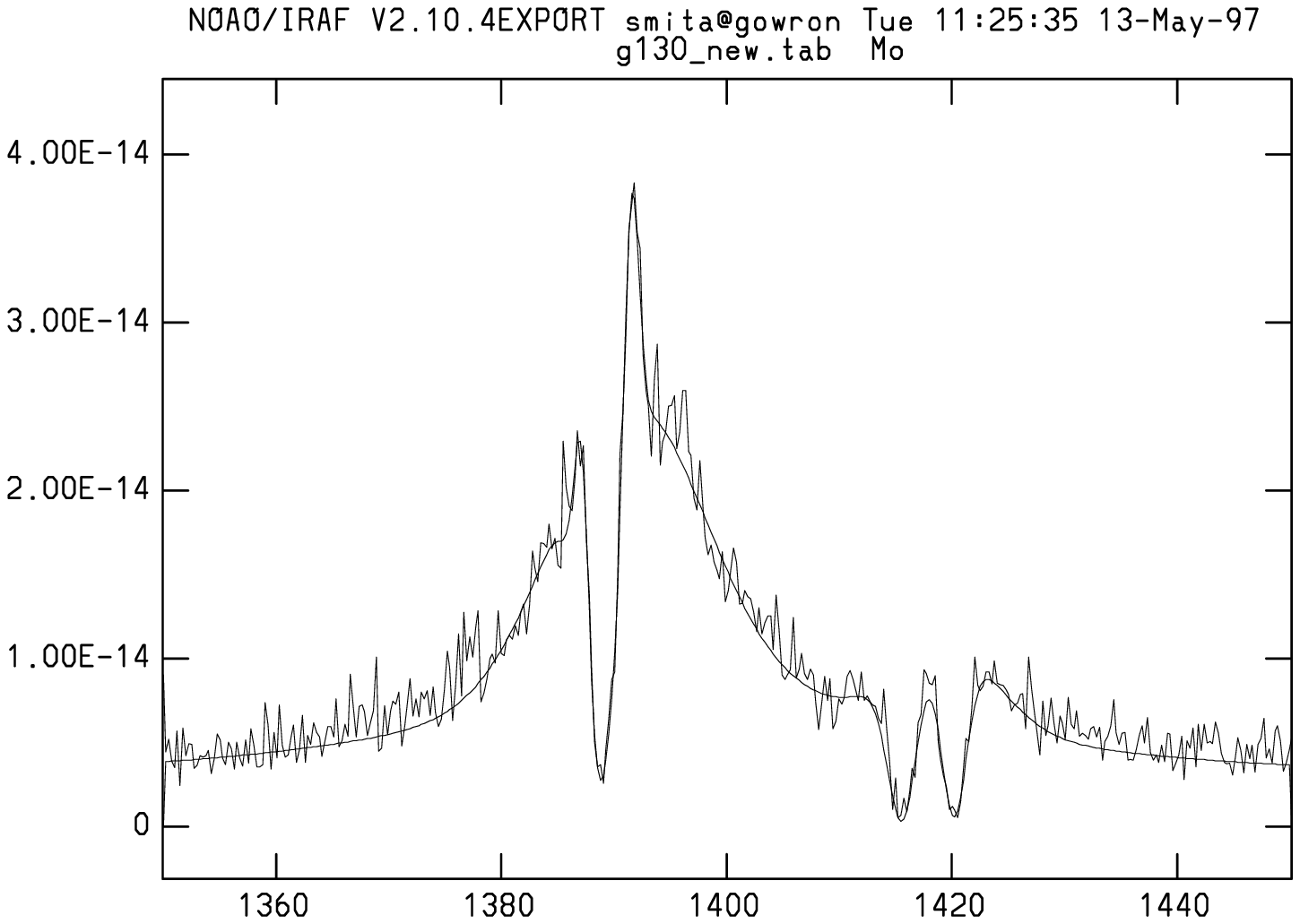,height=3.0in}
\psfig{figure=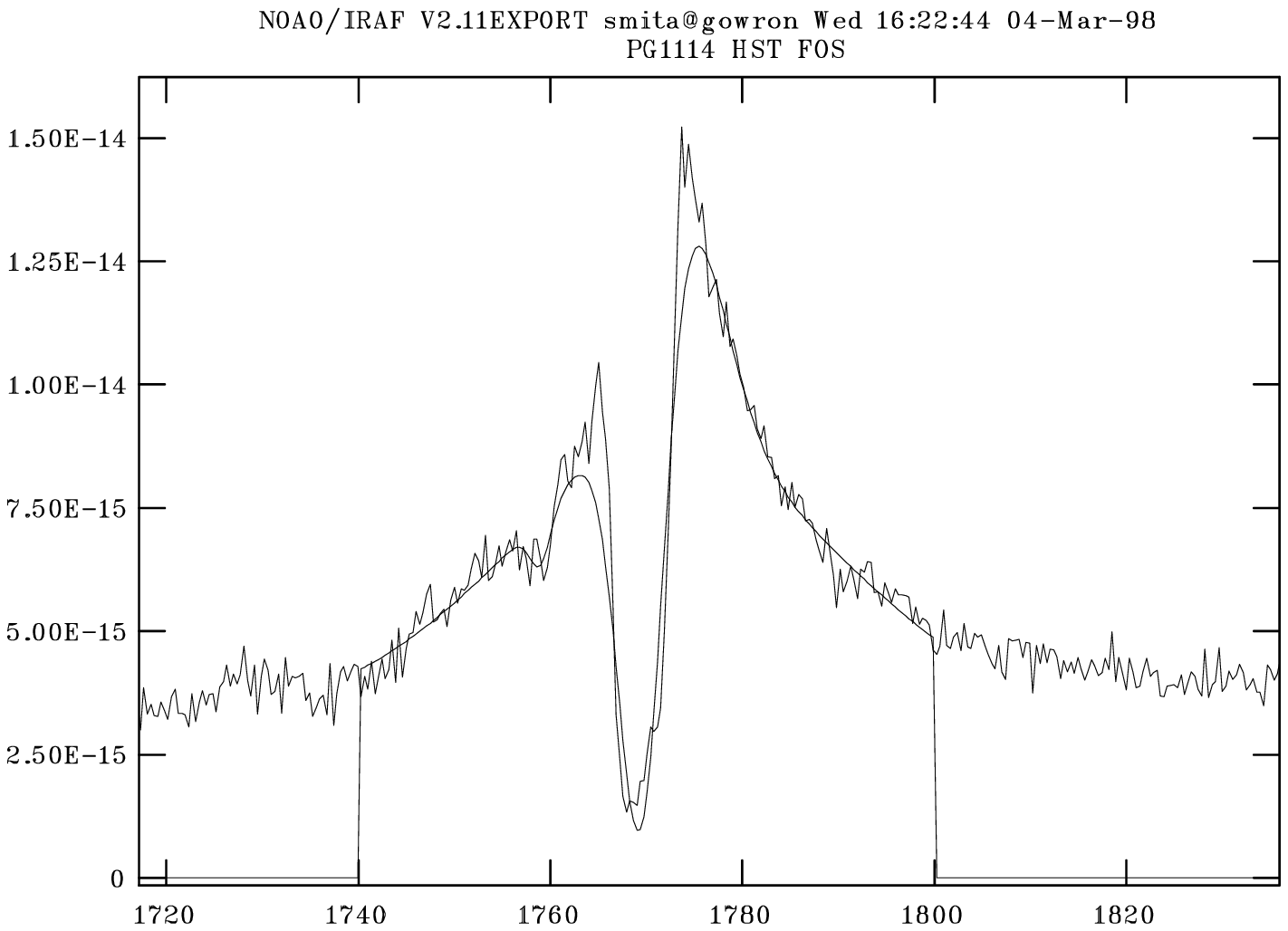,height=3.0in}
\end{figure}

\end{document}